\documentclass[twocolumn,prb,showpacs,floatfix,
superscriptaddress,amsmath,amssymb]{revtex4}

\usepackage{graphicx}
\usepackage{dcolumn}
\usepackage{bm}
\usepackage{color,pstricks}

\begin{document}


\title{Correlations in nano-scale step fluctuations: comparison of simulation and experiments}

\author{F. Szalma}
\email[Corresponding author: ]{szalmaf@physics.umd.edu}
\affiliation{
Department of Physics, University of Maryland, College Park, Maryland 20742-4111}
\author{D. B. Dougherty}
\affiliation{
Department of Physics, University of Maryland, College Park, Maryland 20742-4111}
\affiliation{Department of Chemistry, Surface Science Center, University of Pittsburgh, Pittsburgh, Pennsylvania 15260}
\author{M. Degawa}
\affiliation{
Department of Physics, University of Maryland, College Park, Maryland 20742-4111}
\author{Ellen D. Williams}
\affiliation{
Department of Physics, University of Maryland, College Park, Maryland 20742-4111}
\author{Michael I. Haftel}
\affiliation{Center for Computational Materials Science, Naval Research Laboratory, Washington, DC 20375-5343}
\author{T.L. Einstein}
\email{einstein@umd.edu}
\homepage{http://www2.physics.umd.edu/~einstein}
\affiliation{
Department of Physics, University of Maryland, College Park, Maryland 20742-4111}
\date{\today}

\begin{abstract}
We analyze correlations in step-edge fluctuations using the Bortz-Kalos-Lebowitz kinetic Monte Carlo algorithm, with a 2-parameter expression for energy barriers, and compare with our VT-STM line-scan experiments on spiral steps on Pb(111). The scaling of the correlation times gives a dynamic exponent confirming the expected step-edge-diffusion rate-limiting kinetics both in the MC and in the experiments. We both calculate and measure the temperature dependence of (mass) transport properties via the characteristic hopping times and deduce therefrom the notoriously-elusive effective energy barrier for the edge fluctuations. With a careful analysis we point out the necessity of a more complex model to mimic the kinetics of a Pb(111) surface for certain parameter ranges.
\end{abstract}

\maketitle

\section{Introduction}

Below the roughening temperature a rich variety of structures are observed on crystal surfaces: islands, mounds, shallow and deep pits, self-assembled patterns, step bunches, and spiral steps, as well as such extended structures as quantum wires and arrays of quantum dots.  The evolution of such nanostructures proceeds via mass transport on flat terraces and along the steps separating them.\cite{pimpvillainbook,jeongwilliams99}  
The motion of these structures can now be directly observed and quantitatively determined by monitoring the constituent steps using advanced experimental techniques like variable temperature scanning tunneling microscopy (VT-STM) with near-atomic spatial resolution and with a time resolution fast enough to observe fluctuations on the nanometer scale (but too slow to observe individual atomic movements, as we shall see below). The kinetic parameters relevant to the mass transport mechanisms are typically determined from the spatio-temporal evolution and/or fluctuation of steps. Thus, scrutiny of the relatively simple behavior of steps offers a gateway to understanding the evolution of complex structures and an avenue to test theoretical arguments against the (harsh) reality of experiments.

In this paper we use STM to study thermal fluctuations of spiral steps \cite{madhavspiral05} on the surface of Pb(111) crystallites.\cite{degawacrystallite05} The principal mass transport mechanism on these surfaces is known to be periphery diffusion\cite{spellerPb95,kuipersPb95} near and moderately above room temperature. Less is known, however, about the temperature dependence of the mass transport, with which an effective energy barrier can be associated  in an Arrhenius-type picture due to the activated nature of these processes. This effective energy barrier, which can be measured in experiments, is intimately related to more fundamental parameters characterizing the surface and its steps: the kink energy and the energy barrier in  diffusion. These parameters describe both the equilibrium and out-of-equilibrium behavior of the system, e.g.\ the equilibrium crystal shape, and the linear response of the system to small perturbations, the diffusion coefficient.

We set up a theoretical model and subject it to extensive kinetic Monte Carlo (kMC) 
simulations. In contrast to a step on a vicinal surface, the azimuthal orientation\cite{2006KodambakaKhare} of a spiral step is not well-defined but instead spans all angles.   In the experiments an average is taken over the undefined directions in an imprecise way.  In our model we take a simple average of all azimuthal directions, as one would for an island.\cite{Szalma05} 
These orientation-averaged fluctuations provide the basis for calculating the transport properties and the associated effective energy barriers in this paper.

The paper proceeds as follows.  We first recall the important results of Langevin analysis applied to step fluctuations dominated by atomic motion along the steps.  In the third section we describe the experimental technique for measuring spirals on Pb crystallites and our results.  In Section IV we describe our extensive kMC simulations on a hexagonal lattice using energy barriers determined by a subset of occupied nearest-neighbor sites of each hopping atom with energies computed with a semiempirical scheme optimized for surfaces.  (An appendix expands on the issue of detailed balance in such models.)  Section V presents our kMC results and comparisons with experimental values, followed by a final section of discussion and conclusions.

\section{Background Theory}

Step-edge fluctuations are typically described with stochastic partial differential equations drawn from Langevin formalism.  The characteristic parameters are  equilibrium values determined phenomenologically or from simple models such as a lattice-gas (Ising-like) model of the appropriate symmetry.  We detail here the necessary equations that apply to the fluctuations of steps on the Pb(111) surface. As we will see,  the fluctuations are due to step edge diffusion, which is known to be described by the Cahn-Hilliard equation 
\begin{equation}
\frac{\partial x}{\partial t}=-\frac{\Gamma_h\tilde{\beta}}{k_BT}\frac{\partial^4 x}{\partial y^4}+\eta_h (y,t),\label{eq:CH}
\end{equation}
where 
$x=x(y,t)$ 
is the perpendicular displacement of the step at point $y$ along the step at time $t$. $\Gamma_h$ is the step mobility, which encompasses the friction felt by the step when randomly kicked, $\tilde{\beta}$ is the step stiffness, the inertial parameter which counterweights deviations of the step from its equilibrium shape. 
The conserved thermal noise $\eta_h(y,t)$ satisfies $\langle\eta_h(y,t)\eta_h(y',t')\rangle=-2\Gamma_h\nabla^2\delta(y-y')\delta(t-t')$, where $\langle\rangle$ is the canonical ensemble average.
Here we neglect the fact that the observed objects (the spiral and island) have an intrinsic curvature, but take them as locally straight and apply the same formula as for the straight step. This amounts to leaving out a geometric factor in Eq.~(\ref{eq:CH}) that is close to unity.  

The time correlation function 
$G(t)=\langle [x(0)-x(t)]^2\rangle$ 
is readily obtained \cite{Bartelt92} from the Cahn-Hilliard equation for times shorter than the correlation time of the system 
\begin{equation}
G(t)=\frac{2\Gamma (3/4)}{\pi}\left(\frac{k_BT}{\tilde{\beta}}\right)^{3/4}\Gamma_h^{1/4} t^{1/4}
\label{eq:Gt}
\end{equation}
where the Euler gamma function $\Gamma (3/4) \approx 1.2254$. We denote by $c(T)$ the prefactor of 
$t^{1/4}$; it is the number we typically extract as the endproduct of our experimental measurements. 

The step mobility is directly related to the microscopic time constant $\tau_h$ that indicates the average time between moves at any site (similar to the collision time in transport theory). An equation similar to the Nernst-Einstein relation for the diffusion coefficient describes this relationship $\Gamma_h=\Omega^{5/2}/\tau_h$, where $\Omega$ is the area of the 2d unit cell a Pb atom occupies in the (111) surface. Thus, in experimental analyses based on Eq.~(\ref{eq:Gt}), the time constant is
\begin{equation}
\tau_h=\left(\frac{2\Gamma (3/4)}{\pi \; c(T)}\right)^{4} \left(\frac{k_BT}{\tilde{\beta}}\right)^{3}\Omega^{5/2}.
\label{eq:ctau}
\end{equation}

Though the above equations are all continuum equations, the step stiffness $\tilde{\beta}$ is typically drawn from a simple lattice-gas model.  The transition from continuum to lattice or vice versa implicitly contains the assumption that the steps in question are long enough to be able to form capillary waves with negligible finite-size effects. We calculated the stiffness from the Akutsu formula\cite{akutsuakutsu} for (111) surfaces or its low temperature approximation \cite{loT}
\begin{equation}
\tilde{\beta}=(2kT/3a)\exp[\epsilon_k/kT].
\label{eq:Akut}
\end{equation} 
Here $a$ is the lattice constant of the hexagonal lattice, and $\epsilon_k$ is the kink energy. In the present study we consider only nearest-neighbor (NN) interactions between atoms (in particular when computing hopping barriers) with [attractive] interaction energy $-E_{NN}$, i.e.\ with an effective bond energy $E_{NN}>0$.  In this basic lattice-gas model, there is the simple relation $\epsilon_k=E_{NN}/2$ between these characteristic energies \cite{kinknote}.

The STM measurements are always so-called line-scan measurements: the fluctuations are measured only along a single line perpendicular to the step edge. Since this line is fixed, we cannot gather information about a longer stretch of the edge as one would typically in LEEM measurements and in Monte Carlo simulations.  While this low information density is a severe limitation of these measurements, it is amply compensated by the very high, basically atomic resolution, in contrast to LEEM measurements.

In the MC simulations, since we have data for the whole spatial extension of the step (or island in our case) we can analyze the (spatial) Fourier spectrum of the fluctuations. In this case $x(y,t)$ measures the deviations from the equilibrium 2d crystal shape due to the fluctuations and $x(y,t)=\sum_q x_q(t) \exp [iqy]$, where $q=n 2\pi/L$, $L$ is the circumference of the island; $n=-N/2+1,-N/2+2,...,N/2$, where $N$ is the number of degrees of freedom of the island edge, i.e.\ the number of perimeter atoms. The time correlation function of the Fourier modes \cite{Bartelt92} $G_q(t)=\langle\vert x_q(t)-x_q(0)\vert^2\rangle$ exponentially approaches the equilibrium fluctuation (roughness) of the step
\begin{equation}
G_q(t)=\frac{2k_BT}{L\tilde{\beta}q^2}\left(1-{\rm e}^{-\vert t\vert/\tau(q)}\right).
\end{equation}
The correlation time $\tau(q)$ scales with the wave number as 
\begin{equation}
\tau(q)=\frac{k_BT}{\tilde{\beta}}\frac{\tau_h}{\Omega^{5/2}q^z}
\label{eq:tauq} 
\end{equation}
with the dynamical exponent $z=4$ in the periphery diffusion case, where $\tau_h$ is the same hopping time as above.

Since the kinetic processes are activated processes on the surface, the hopping rate
\begin{equation}
\frac{1}{\tau_h}=\nu {\rm e}^{-E_h/k_BT}
\end{equation}
can be characterized by an effective hopping barrier $E_h$ and an attempt frequency $\nu$. The hopping barrier combines the kink energy $\epsilon_k$ responsible for the kink density along the step and the (edge-)diffusion barrier $E_d$. For any particular crystallographic direction, this combination comes with definite prefactors. However, because we average over all directions, this decomposition becomes irrelevant; therefore, we directly measure the hopping barrier both in the experiment and the MC simulations. We expect the attempt frequency to be of order the Debye frequency since it is mostly the lattice vibrations that give rise to adatom hops on the surface.

\section{Experimental Methods and Results}

Supported 3D Pb crystallites were prepared on Ru(0001) by dewetting a continuous film (10-30 nm nominal thickness) under UHV conditions.  The film was prepared by depositing Pb from a heated alumina tube onto a room-temperature Ru(0001) substrate.  By heating the sample to the melting point of Pb (600 K) for several minutes, the continuous film formed liquid beads that were then quenched to form the solid crystallites.  Crystallites formed in this way have (111) facets parallel to the Ru substrate plane that can be as large as 1 $\mu$m in diameter.  {

\begin{figure}[t]
\includegraphics[angle=-0,width=.9\columnwidth]{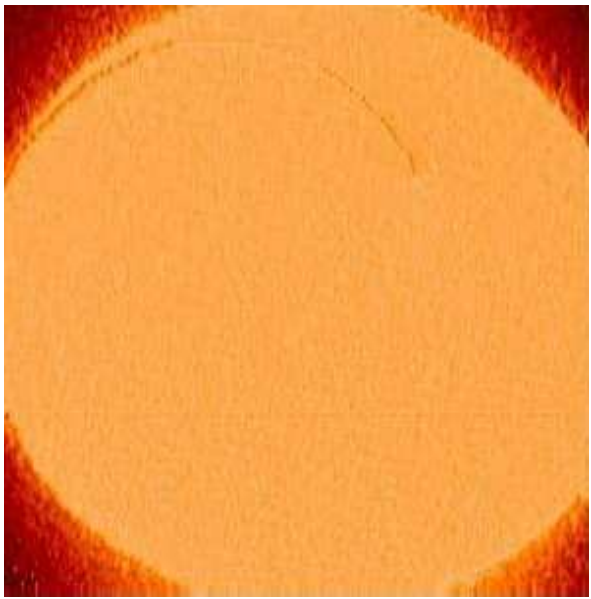}
\caption{\label{fig:spiral} (Color online) 1000nm x 1000nm STM image of a spiral dislocation on top of a Pb(111) crystallite. The STM line-scan measurements were performed perpendicular to the step edge, typically 5--50nm from the dislocation core.}
\end{figure}

STM measurements were made with a commercial instrument (Omicron VT-STM) at room temperature and above. The sample was heated by a PBN heater in the sample holder (Omicron), and the temperature was calibrated using a manufacturer-supplied curve that was checked with infrared pyrometry for $T >$ 520~K. Step fluctuations were monitored on (well isolated) steps using the line-scan method of imaging: the STM tip is fixed over one point along the step edge and scanned perpendicularly for 20 - 120 seconds.  The step position from such an image could be extracted and used to compute $G(t)$, as described in Section II.

\begin{figure}[b]
\includegraphics[angle=-0,width=.9\columnwidth]{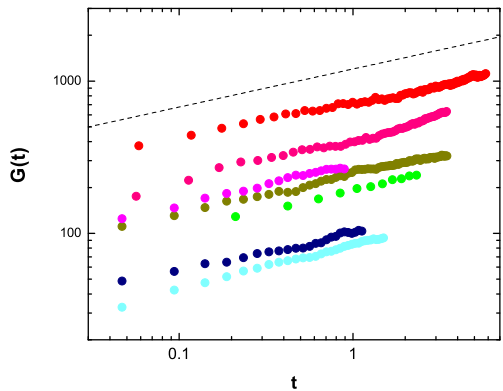}
\caption{(Color online) Log-log plot of equilibrium correlation functions (measured in \AA$^2$) vs. time (sec.). Measured in the configuration depicted in Fig.~\ref{fig:spiral}, the data sets are  for temperatures 300K, 320K, 330K, 340K, 350K, 370K, 390K, from bottom to top, respectively. The slope of the curves give the growth exponent and determine the rate limiting process responsible for the fluctuation of the steps. The thick dashed line corresponds to an exponent 1/4.  Evidently the data for the temperatures depicted in this plot are essentially consistent with this value (see text), suggesting that the rate limiting process is step-edge diffusion.}
\label{fig:expgvst}
\end{figure}

{

We make our measurements in the 300K-400K temperature range. The
experimental situation strongly changes in this range. At high
temperatures ($>350$K), kinetics are fast and small structures
accidentally created on the facet such as islands, voids or mounds,
decay away fairly fast.  Isolated steps are only obtained when there is
a screw dislocation on the facet as shown in Fig.~\ref{fig:spiral}. The
spiral step terminating at the screw dislocation continues to unwind
until it reaches equilibrium with the crystallite.  For temperatures
$<350$K, this unwinding process becomes slow, so that it is more
difficult to determine how close the step is to equilibrium.  As for room
temperature, small structures do not decay easily.  Thus for room
temperature experiments, in addition to spiral
steps, we consider steps crossing in-between small structures on the
facet, treating them as isolated steps which are assumed to be in local
equilibrium.  

We have made two sets of measurements, one with less temperature
resolution, in which we explore the coarse structure of the temperature
dependence, and another with higher resolution, in which we pay more
attention to the details. We applied different analyses to the two sets;
there are also differences in the experimental environment. The effects
of these differences will be discussed later. 

Experimentally determined temporal correlation functions (from the
higher-temperature resolution measurements) for different temperatures
are shown as a log-log plot in Fig.~\ref{fig:expgvst}.  The points are
the results of averaging several (5-15) experimental $G(t)$ curves
obtained from displacement-vs.-time images taken over typically a few
minutes at the same point of the step edge for each temperature.  For
screw dislocations most measurements were done relatively close to the
dislocation core (within 5-50nm, which is also far enough from the core, so that the screw step has reached its full value), although we do not see a significant difference in the trial measurements farther from the core.

The $G(t)$ data are linear on log-log plots, indicating power-law growth
of $G(t)$; their slopes are consistent with the $t^{1/4}$ behavior
(indicated by the slope of the dashed line in Fig.~\ref{fig:expgvst})
expected for step-edge diffusion (cf.\ Eq.~(\ref{eq:Gt})). 
Specifically, their exponent values are close to 1/4, ranging from 0.20
to 0.35 and clustering around 0.27. Consistent with Eq.~(\ref{eq:Gt}),
the magnitude of $G(t)$ (i.e.\ the prefactor $c(T)$) increases
dramatically as temperature increases.  

From a fit of the experimental curves like the ones in
Fig.~\ref{fig:expgvst}, using Eq.~(\ref{eq:Gt}), $c(T)$ is obtained.  In
Fig.~\ref{fig:expprefact} we show the experimental prefactors as a
function of temperature. Each data point is obtained from an average of
all $G(t)$ curves for each measured step position. The set of
measurements with little temperature resolution is indicated by
triangles, while the ones giving more detail about the temperature
dependence are denoted by squares. The latter set consistently gives
somewhat higher prefactor values probably due to the different image
processing and analysis used.

{ 

\begin{figure}[b]
\includegraphics[angle=0,width=.9\columnwidth,clip=true]{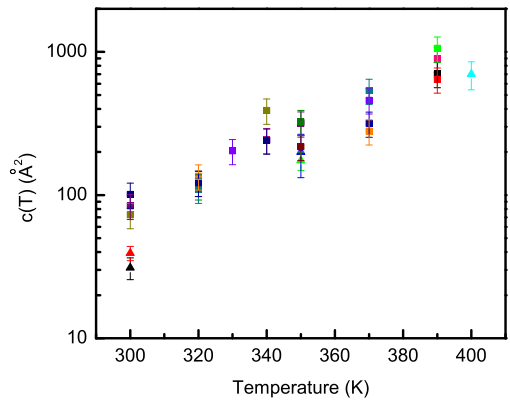}
\caption{\label{fig:expprefact}  (Color online) Prefactors $c(T)$ of the experimentally
measured correlation functions vs. temperature, calculated based on
Eq.~\ref{eq:Gt} assuming step-edge diffusion as the rate limiting
process for the fluctuations of the steps. For each temperature, the
different data points correspond to the averaged prefactor at each
measured step position. The triangles and squares indicate two different
sets of measurements (see text).}
\end{figure}

We use the prefactors to extract the step-edge diffusion time constant
$\tau_h$ with the aid of Eq.~\ref{eq:ctau}.  It is found to decrease
monotonically from 50$\mu$s to 0.073ns as temperature increases from
300K to 400K.  Due to the fourth power of $c(T)$ in Eq.~(\ref{eq:ctau}),
the error in the determination of $\tau_h$ can  be very significant.  In
later sections this experimental uncertainty will be discussed more
fully.  These findings for $\tau_h$ are discussed in
Sec.~\ref{sec:results} and depicted in Fig.~\ref{fig:tauvsT} along with
the MC results.

\section{Kinetic Monte Carlo}

In order to investigate correlations in a ``clean" environment, we perform a kinetic Monte Carlo simulation of a hexagonal lattice gas. Simulational data can provide corroboration of our fundamental picture of what transpires during the experiments and can also help in their analysis. It has the advantage that far more data can be collected, allowing for an even more precise analysis, and it also has spatial resolution (\textit{along} step edges) which would be difficult to get in today's STM measurements, as will become more evident shortly. While these measurements can acquire fluctuation data in a line scan, which provides us with fluctuation in time at a particular point along the step edge, the kMC provides us with instantaneous snapshots of a spatially extended region of the step edge (again along the step), in our case snapshots of the whole island edge. Data similar to what we can get in the kMC simulation can be obtained by LEEM,\cite{BarteltTromp} but in that case the spatial resolution both along and perpendicular to the step is less than the resolution of the simulation (or the resolution of the STM perpendicular to the step edge). 
Thus, both experimental methods have their advantages and disadvantages over the other while the advantages of the MC mentioned above is diminished by the fact that it inevitably contains simplifications in the geometry of the system and the energy landscape of the surface determining the kinetic barriers for the hops of the adatoms and also for the configuration of the hops. (Only NN hops of single adatoms are considered in the simulations.)

In considering the (111) surface of Pb to be a 2D, hexagonal net with 6-fold symmetry, we disregard the $ABC$ stacking structure leading to only 3-fold surface symmetry. The adatoms on the surface are put into a circular ``container" with an impermeable wall. The adatom gas in the container is a highly supersaturated (but low-density) lattice gas at the temperatures of present concern; hence, most of the atoms in the 2D gas condense into a big island close to the middle of the container, surrounded by detached adatoms moving around the terrace between the island and the container wall. This very rarified remaining gas of mobile adatoms and edgeatoms (adatoms sliding along the island edge) gives rise to the fluctuations of the island edge. The island is of monatomic height; we completely neglect the possibility of an adatom hopping onto the top of the island.

We use the Bortz-Kalos-Lebowitz\cite{Bortz75} (BKL) continuous-time MC algorithm in our simulations since it is well suited to the low temperatures of interest here. The critical temperature in the simulations is $T_c=$~1400K, while the typical temperatures in question are below 400K. The rejection-free feature of the BKL algorithm greatly improves the efficiency of the simulations, and efficiency is further enhanced by the $n$-fold-way method, since $n$ in our case is quite low ($n=5$) due to the small number of different energy barriers.

We evaluate energy barriers using semiempirical surface embedded atom method\cite{SEAM} (SEAM) calculations using Ercolessi's glue potentials\cite{Erco} for the Pb(111) surface (fourth column of Table \ref{tb:energybarriers}).
Our BKL simulation of the lattice gas 
expresses these barriers in terms of two characteristic energies. One of these parameters is the barrier for an adatom to hop on the terrace when it has no lateral neighbors, i.e., surface diffusion; a direct computation of this energy barrier with SEAM gives $E_{TD}$=70 meV (see Table \ref{tb:energybarriers}).  The other parameter accounts for the possibility that the adatom has lateral neighbors with which ``bonds" [direct and indirect] are broken when the adatom hops.  We consider only NN bonds as depicted in Fig.~\ref{fig:nnconfig}.  

In our first parametrization, we adopt a scheme we call ``break-3" in which the energy barrier of a hop depends on how many sites are occupied among the sites ``behind" a hop, labelled 0, 1, or 1$^\prime$; they each contribute $E_{\rm NN}$=130 meV, which is the other characteristic barrier of the lattice gas. $E_{\rm NN}$ is obtained via trial and error comparison with EAM barriers; see Table \ref{tb:energybarriers} and the next paragraph.
\begin{figure}[b]
\includegraphics[angle=-90,width=.7\columnwidth]{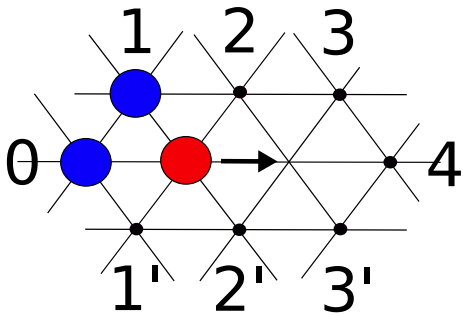}
\caption{\label{fig:nnconfig}  (Color online) Sample configuration on triangular lattice, indicating our nomenclature for the nearest-neighbor sites of the adatom before and after the move. In this case the move is labelled 01 since site 0 and 1 are occupied among the eight NN sites surrounding the two sites involved in the hop. 
See Table \ref{tb:energybarriers} for the energy barriers of the various hop configurations.}
\end{figure} 
We neglect any effect of the other nearest-neighbor sites to the two sites involved in the hop (2, 2$^\prime$, 3, 3$^\prime$, and 4), which we expect to be smaller. The actual energy barrier $E_b$ for such a hop is the sum of the terrace diffusion barrier and the contributions of the bond breakings
\begin{equation}
E_b=E_{TD}+mE_{\rm NN},\label{eq:Eb}
\end{equation} where $m$ is the number of occupied ``behind" sites with which a bond is broken, and it ranges from 0 to 3. 

On closer inspection the energy barriers computed as above with SEAM (or any form of EAM) seem not to satisfy detailed balance for more complex processes such as the one depicted in Fig.~\ref{fig:microstates} and analyzed in the Appendix. The reason is that these processes involve next-nearest-neighbor (NNN) and possible further-neighbor (or multi-neighbor\cite{TSHGTE}) interactions; thus, EAM only satisfies detailed balance if these are also taken into account, so that the hopping process is considered in its full complexity. However, to allow for long kMC runs and transparent results, we only want to capture the main hopping mechanisms relevant to the material; the full details of the energy landscape are an uninteresting  complication and distraction. We use the EAM barriers as a guide to deduce $E_{TD}$ and $E_{NN}$ and in this way simplify them into classes such as in the break-3 scheme just described (column two). This scheme along with the other simplified schemes described below satisfies detailed balance as shown in the Appendix.

Another such classification is the break-five scheme in column three, which involves sites 0, 1, 1$^\prime$, 2, and 2$^\prime$. Each of these sites, if occupied, contributes $E_{NN}$ to the hopping barrier, so that in Eq.~(\ref{eq:Eb}) $m$ ranges from 0 to 5.  
We expect that the break-3 and break-5 parameterizations bracket the actual dependence of the barrier on lateral occupation. 
\begin{table}
\begin{tabular}{l||r|r|r|r|r}
Config & $E_b^{bb3}$  &
         $E_b^{bb5}$ & 
      $E_b^{\rm EAM}$  & 
          $\Delta E_b^{\rm EAM}$  & 
         $E_b^{\rm Kaw}$  \\

\hline
TD                        &   70 &    70 &  70  &   0   &  70 \\
0                          & 200 &  200 & 192 & 116 & 200 \\
01                        & 330 &  330 & 260 & 225 & 330 \\
12                        & 200 &  330 & 237 & 147 & 200 \\
012                      & 330 &  460 & 359 & 269 & 330 \\
011$^\prime$    & 460 &  460 & 467 & 386 & 460 \\
123                      & 200 &  330 & 108 &   0    &  70 \\
011$^\prime$2  & 460 &  590 & 598 & 469 & 460 \\
1234                    & 200 &  330 & 141 &-162 &  70 \\
23                        &   70  &  200 &  86  &-147 &  70 \\
22$^\prime$      &   70  &  330 &130  &   0   &  70 \\
11                        &  330 &  330 &312 & 235 & 330\\
023                      &  200 &  200 &135 & -32  & 70 \\
\end{tabular}
\caption[shrtab]{\label{tb:energybarriers} Tabulation of energy barriers (all in meV) for selected hopping processes. See Fig.\ \ref{fig:nnconfig} and text for explanation of the configurations. The 3 bond-breaking (bb3) scheme barriers, 5 bond-breaking scheme (bb5), the EAM energy barriers, the EAM energy difference of two states,  and the Kawasaki-type barriers are listed, respectively. We use the bb3  scheme in the kMC simulations in this paper.}
\end{table}

The break-3 classification is evidently quite a good approximation to the EAM in the 0, 12, 012, 011$^\prime$, 23, 11$^\prime$ cases. The break-5 is good in the 0, 011$^\prime$, 11$^\prime$ and especially the 011$^\prime$2 cases but in several other cases significantly worse than the break-3 scheme, e.g.\ 123, 1234, 23, and 22$^\prime$ cases. 

For the sake of comparison, we also list the Kawasaki-type energy barriers\cite{type} (column six) that are usually calculated based on how many bonds have been broken and how many new ones have been formed in a move, plus the usual terrace diffusion barrier. This scheme is often used in surface simulations. Its philosophy is markedly different from the previous two bond-breaking schemes because it takes into account the newly formed bonds, i.e.\ the energy of the final state. The Kawasaki scheme is not regarded the best scheme for surface simulations,\cite{NewBark99} but in our case Table \ref{tb:energybarriers} shows it to be quite close to the EAM. 

\section{Results} \label{sec:results}

To check the precision of the kMC procedure, we compared the equilibrium island shapes to analytical results by Zia\cite{zia} for a triangular lattice.\cite{TSHGTE}
In Fig.\ \ref{fig:eqshape} these shapes are depicted for T=250K and show excellent agreement. The slight difference between the MC shapes and the 
exact shape may be due to the fact that the shapes are always measured from the instantaneous center of mass, which may give rise to more rounded shapes in the MC. 
Furthermore, since the islands are not large compared to the lattice spacing $a$, there might well be finite size effects (which we have not attempted to analyze here), but this effect should also appear in the experimental studies of nano-islands having the same size as in the MC simulations.

\begin{figure}[b]
\includegraphics[width=.9\columnwidth]{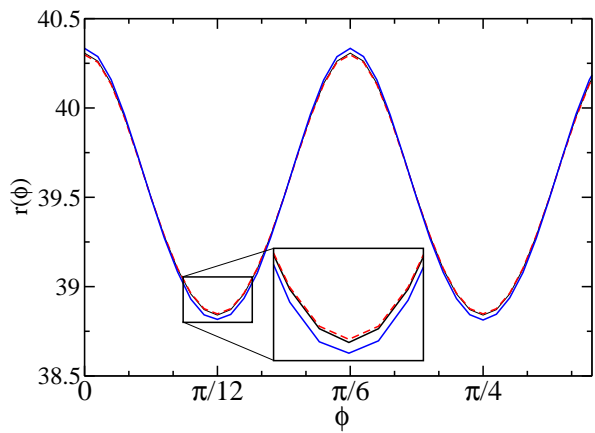}
\caption{\label{fig:eqshape} (Color online) The $r$ vs.\ $\phi$\ plot of the equilibrium crystal shape at T=250K, for radii r=20a, r=40a, and the exact shape in the continuum limit, denoted by dashed (red), thin solid (black), and thick solid (blue) lines, respectively. The graphs are appropriately rescaled so that the islands have the same size to make shape comparison possible for islands of different physical sizes. (This transformation is a mathematical similarity transformation which does not change the shape of the islands, only their size.) The inset shows the minimum of the curves magnified for better comparison.}
\end{figure}

From the MC data we calculate the line tension and the step stiffness in the two principal directions of the 6-fold symmetric triangular lattice making 30$^\circ$ angle. They compare favorably (see Table \ref{tb:beta}) with exact results and with measured values (of line tension) for Pb(111),\cite{bombis-bonzel} indicating that
the Ising model and the corresponding lattice gas model
with the chosen energy barriers describe these structures well. The deviations of the KMC from the exact data and the measured ones basically show the error bars and the limit of applicability of the model and of the simulations. 

\begin{table}[t]

\begin{tabular}{l||r|r|r|r}
T (K) & \shortstack{$\beta$(0$^\circ$) \\(meV/\AA)\\ exact} & 
          \shortstack{$\beta$(0$^\circ$) \\(meV/\AA)\\ KMC} & 
          \shortstack{$\tilde{\beta}(0^\circ)/\tilde{\beta}(30^\circ)$\\ exact} &
          \shortstack{$\tilde{\beta}(0^\circ)/\tilde{\beta}(30^\circ)$\\ KMC}\\
\hline
250 &  36.55 &  34.1 & 3.99 & 4.09 \\
300 &  35.98 &  34.2 & 2.49 & 2.37 \\
350 &  35.28 &  34.0 & 1.83 & 1.80 \\
400 &  34.34 &  33.8 & 1.49 & 1.49 \\
\end{tabular}
\caption[shrtab]{Tabulation of the line tension and the 
relative stiffness to test the KMC simulation on the 
exact lattice gas results by Akutsu and Akutsu.\cite{akutsuakutsu} 
In comparison, experiment\cite{bombis-bonzel} finds 29.65meV/\AA\ for 
323K, 28.65meV/\AA\ for 353K, 27.2meV/\AA\ for 393K  for the line tension (i.e., the average of that of A and B-type steps, since we neglect their difference in the simulations).}
\label{tb:beta}
\end{table}

In our KMC simulations the step fluctuations are driven by step-edge diffusion throughout the studied temperature range. The scaling of the 
correlation time $\tau(q)$ with wavenumber gives a dynamic 
exponent $z=4$ characteristic of these processes (see Fig.\ \ref{fig:tauvsq}). At higher temperatures we might expect a crossover to other processes like terrace diffusion or attachment-detachment with $z=3$ or 2 dynamic exponents, respectively. These processes could show up due to the increased roughness of steps at higher temperatures, which give rise to sites in the step edge that are bound to the step by only two bonds instead of the more typical three or four, driving the system into the attachment-detachment rate limiting regime (see {Ref.[\onlinecite{2001SzalmaSelkeFisher},\onlinecite{98Ihleetal},\onlinecite{2005SzalmaWilliams}]}).  But this scenario is confirmed neither by the experiments\cite{DBD} nor the MC simulations here. 

\begin{figure}[h]
\includegraphics[width=\columnwidth]{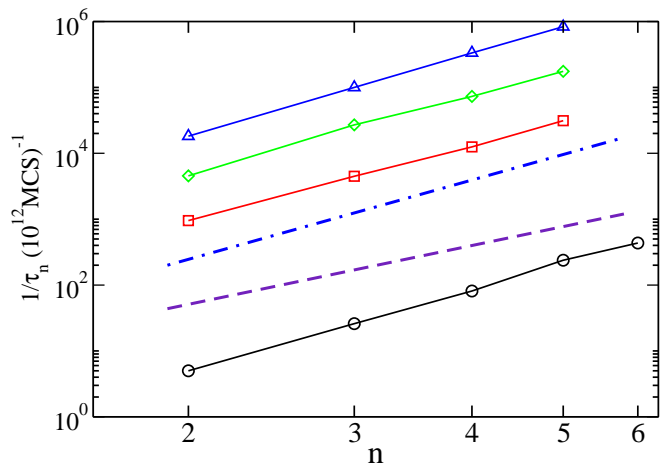}
\caption{\label{fig:tauvsq} (Color online) Dynamical scaling for the KMC data. Scaling of the relaxation time $\tau_n$ versus the wavenumber $n$ for $T=$250K (circle), 300K (square), 350K (diamond), 400K (triangle). For the lowest temperature $r=40a$, in the other cases $r=20a$. In all four cases the dynamical exponents, given by the slopes of the curves, suggest step-edge diffusion limited kinetics, see Table \ref{tb:tauh}.  The straight dotted-dash and the dashed lines are guides (at arbitrary ordinate) to show the expected slopes for step-edge diffusion ($z$=4) and terrace diffusion ($z$=3), respectively.}
\end{figure}

The hopping time $\tau_h$ characterizes the average time that elapses between successive ``events" at a site, viz.\ the arrival or departure of an adatom. This change of occupation can be part of either a step-edge diffusion process (which is the more typical here) or an attachment-detachment process associated with terrace transport. Using the exact step stiffnesses and the correlation times $\tau(q)$ of the shortest wavelengths that we measured in the MC simulations in Fig.\ \ref{fig:tauvsq}, we calculate the more fundamental time constant $\tau_h$ for each temperature from the correlation time of the Fourier modes via Eq.\ \ref{eq:tauq}.

\begin{figure}[b]
\includegraphics[width=.9\columnwidth]{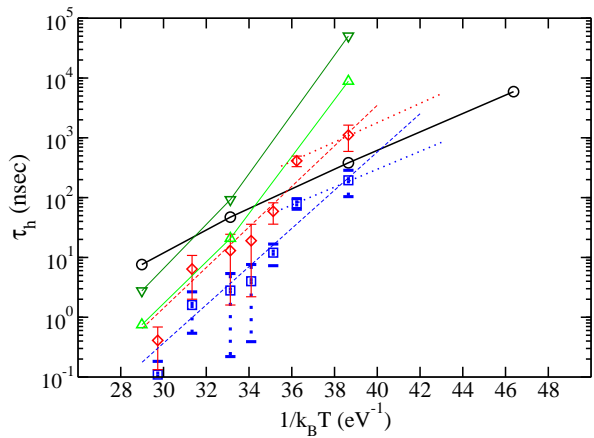}
\caption{(Color online) Arrhenius plot of hopping times. Two sets of experimental results (triangles and squares) are plotted with two different kink energies, $\epsilon_k$=50meV (downward pointing triangles, and diamonds) and 65meV (upward pointing triangles, and squares) (see Eq. (\ref{eq:tauq})). The KMC data are depicted in circles with a thick solid-line as a guide to the eye. The error bars are indicated on the experimental data along with a linear fit (dashed lines) on the higher temperature resolution data.  The dotted lines indicate how we speculate these data might continue if the experiments had gone below room temperature. The slopes of the fits (dashed lines) to all data points except the highest temperature ones show the energy barriers to be apparently higher than that of the MC (see text for discussion, and also Table \ref{tb:energies}). }
\label{fig:tauvsT}
\end{figure}

As shown in Fig.~\ref{fig:tauvsT}, the Arrhenius plot of the Monte Carlo $\tau_h$ gives the effective energy barrier $E_h$=380 meV and a frequency $\nu_0=7.0\cdot 10^{12}$ Hz, which is in the range of the Debye frequency. We also plot the two sets of experimental $\tau_h$, in both cases using two different kink energies, 50 meV and 65 meV, in $\tilde{\beta}$ in Eq.~\ref{eq:Akut}. Here 50 meV is the average measured kink energy of the A- and B-type steps, while 65 meV is the kink energy we get in our EAM calculations. 
The two experimental data sets with both kink energies basically provide us with the same effective energy barrier. The two sets are offset by an order of magnitude which gives different prefactors, $\nu$. This difference might be due to the earlier mentioned different analysis used in both sets of measurements and shows the susceptibility of the prefactors and the robustness of the energy barrier.
{

Though the two data sets seemingly give similar results for the
effective energy barrier, the data set with higher temperature
resolution reveal more subtleties, so we analyze this data set here. For
low temperatures the experiments apparently bracket the MC  with the
same energy barrier, while for higher temperatures this experimental
data set tends to bend below the MC, suggesting a higher effective
barrier. If we crudely fit the experimental data without using the MC as
a guide, we get 780 meV and 735 meV, using the two kink energies,
respectively. These effective barriers for experiments at higher
temperature seem to differ greatly from the nice scaling of the various
energies for comparable materials such as Pt, Ag and Cu (see Table
\ref{tb:energies} and Sec.\ VI), suggesting the conclusion that a
phenomenon that is not relevant on the other surfaces --- also not
captured by the (nearest-neighbor single-atom-hop) MC algorithm ---
comes into play on the Pb(111) surface. We elaborate on this idea in the
following Section. {


\begin{table}[t]
\begin{tabular}{l||c|r|r|r|r|c}
T (K) & $z$ & $L$ (\AA)& $n$ & $\lambda$(\AA) & $\tau(\lambda)$ &
$\tau_h$ \\
\hline
250 & 3.98 &  879.2 &  6 & 146.5 & 1.26ms   & 5.91$\mu$s\\
300 & 3.81 &  439.6 &  5 & 87.9  & 17.5$\mu$s   & 0.382$\mu$s\\
350 & 4.02 &  439.6 &  5 & 87.9  & 3.1$\mu$s & 47.1ns   \\
400 & 4.17 &  439.6 &  5 & 87.9  & 0.66$\mu$s & 7.59ns   \\
\end{tabular}
\caption[shrtab]{Tabulation, for the four temperatures used in the KMC
simulations, of the dynamical exponent $z$, system size (viz.,
circumference) $L$ , the wavenumber $n$ and wavelength $\lambda$ used in
calculating the correlation times, the correlation time $\tau(\lambda)$,
and the hopping time $\tau_n$. Due to the activated nature of the
processes, the hopping times change by a significant factor over the
given temperature range. } 
\label{tb:tauh}
\end{table}

\section{Discussion and Summary}

{

The diffusion barrier along a {\it straight} step is usually deduced
from measurements \cite{01giesenrev} to characterize step-edge diffusion
kinetics. Its measurement is based on considerable amount of assumptions
such as low edge atom density, which also entails slightly fluctuating
steps. Both the fluctuation and the geometric kink density is considered
to be low. With these assumptions the correlation function 
$G(t)=(2/\pi)\Gamma(3/4)P_k^{3/4}(c_{st}D_{st})^{1/4}t^{1/4}$  where
$P_k\approx2\exp[-\epsilon_k/k_BT]$ is the kink density,
$c_{st}=\exp[-E_{st}/k_BT]$ the adatom density at the step edge, and
$D_{st}=D_0 \exp[-E_d^k/k_BT]$ is the tracer diffusion along a kinked
step edge. A common simple assumption is that the adatom creation energy
at the edge is $E_{st}=2\epsilon_k$, and the kinked step diffusion
energy barrier is related to the straight step-edge diffusion energy
barrier via $E_d^k=\epsilon_k+E_d$.  $D_0$ is the prefactor of the
diffusion coefficient. Since the kink energy $\epsilon_k$ is usually
known from other types of measurements, the diffusion barrier $E_d$ can
be determined by analyzing the $G(t)$ correlation function since
$4E_t=3\epsilon_k+E_{st}+E_d^k=6\epsilon_k+E_d$, where $E_t$ is the
measured activation barrier of $G(t)\propto \exp[-E_t/k_B T] t^{1/4}$.
In Table \ref{tb:energies} these diffusion barriers are tabulated for
Pt, Cu, and Ag, for which these energies are known, along with the kink
energy and the bulk cohesion energy for the sake of comparison in trends.

\begin{table}[h]
\begin{tabular}{l||c|c|c|c|c}
 & $E_{coh}$ & $\epsilon_k^{Exp}$  & $E_d^{Exp}$ &  $\epsilon_k^{Th}$  &
$E_d^{Th}$ \\
\hline\hline
Pt  & 5.84\footnote[1]{Ref.[\onlinecite{kittel}]} & 
167\footnote[2]{Ref.[\onlinecite{giesenPtfluct}]} & 
1000\footnotemark[2] & 161(A)
178(B)\footnote[3]{Ref.[\onlinecite{94khare}]} & 840(A)
900(B)\footnote[4]{Ref.[\onlinecite{99afeibelman}]} \\ \hline
\raisebox{1.3ex}[0ex]{Cu} & \raisebox{1.3ex}[0ex]{3.49\footnotemark[1]}
&  \shortstack[c]{128\footnote[5]{Ref.[\onlinecite{95giesenss}]} \\
113\footnote[6]{Ref.[\onlinecite{01giesenetal}]}} & 
\raisebox{1.3ex}[0ex]{320\footnotemark[5]} & \raisebox{1.3ex}[0ex]{90(A)
120(B)\footnote[7]{Ref.[\onlinecite{99bfeibelman}]}} &
\shortstack{228\footnote[8]{Ref.[\onlinecite{94stoltze}]} \\
290\footnote[9]{Ref.[\onlinecite{95karimi}]}}  \\ \hline
Ag & 2.95\footnotemark[1] & 
101\footnote[10]{Ref.[\onlinecite{01giesenrev}]} & 
0$\pm$100\footnote[10]{Ref.[\onlinecite{01giesenrev}]} &
74\footnote[11]{Ref.[\onlinecite{99vitos}]} & 220\footnotemark[8]  \\ \hline
\raisebox{1.3ex}[0ex]{Pb} & \raisebox{1.3ex}[0ex]{2.03\footnotemark[1]}
&   \shortstack{40(A) 60.3(B)\footnote[12]{Ref.[\onlinecite{02emundts}]}
\\ 61(A) 87(B)\footnote[13]{Ref.[\onlinecite{99arenhold}]}} & &
\raisebox{1.3ex}[0ex]{41(A)
60(B)\footnote[14]{Ref.[\onlinecite{00feibelman}]}} & 
\raisebox{1.3ex}[0ex]{\bf185}  \\
\end{tabular}
\caption[shrtab]{Tabulation of the cohesion energy $E_{coh}$ [eV/atom],
and for the (111) surface the kink energy $\epsilon_k$ [meV], and  the
diffusion barrier along straight steps $E_d$ [meV]. For these four
materials there is a clear trend both in the experimental (Exp) and
theoretical (Th) values of these energies, where Pb proves to be the
``softest". For $E_d$ for Pt, Cu, Ag the theoretical values are from
embedded atom (EAM) and ab initio (density functional) calculations. For
$E_d$ for Pb we give our KMC result, which is corroborated by our
experiments at lower temperatures; this value excellently follows the
trend of the other columns of the table.}
\label{tb:energies}
\end{table}

In our case, we extract part of the above $G(t)$, namely
$\tau_h=1/c_{st}D_{st}$, and in the above approximation our hopping
energy is $E_h=3\epsilon_k+E_d$. From the MC simulation, for the
diffusion energy we deduce $E_d$=185meV, which is close to the value
$E_d$=200meV predicted from our bond-breaking scheme with our
semiempirical input energies. The under-10$\%$ difference shows how
effective the assumptions are in the above argument. In spite of the
fact that in the MC simulation we consider a small island with very
large curvature and obviously many geometric kinks, the deduced
diffusion barrier along the step edge is very close to the prediction
based on a straight unkinked edge. 

For the lower end of the experimentally measurable range (viz.\ room temperature), our simulational results agree with these experiments. It is unfortunate that going lower temperatures in the experiments was not possible because of the very slow equilibration of the samples.

In similar analyses of the higher-temperature experiments, this barrier
increases by 400meV.  Such a barrier is twice as high as for Cu, closer
to Pt. Since this is strongly counter to the trend in the cohesion
energy and the kink energy, we should consider other
phenomena that are related to the special spiral geometry in our
measurements or ones not present for the other materials.
These higher energy barriers for higher $T$ may be an indication of
different regions of transport mechanisms on Pb(111). The downward bend
at higher temperatures might be an indication of other transport
mechanisms setting in that are excluded in our MC algorithm, such as
concerted moves of dimers or other multi-mers,\cite{rahman97} or one
might think of corner rounding\cite{2001ZhongLagally,2006ShimAmar} having a significant effect, but not
included in our MC as it requires next nearest neighbor hops.

The effective energy barrier may well depend on the actual geometry of
the sample. Different curvatures might lead to different energy barriers
due to change in the density of geometric kink sites. It may also depend
on whether the step edge is compact or more fractal-like since the
energy barrier contains a weighted combination of the few typical
bond-breaking barriers where the 
weights depend on the average local geometry around hopping adatoms on
the step edge. This argument also implies that these effective energy
barriers may even  depend on temperature since the local geometry may
change since the roughness increases with temperature. However, these
geometric arguments do not seem to be corroborated by the MC
simulations, as we have just seen.

The larger scatter in the measurement data both in this work and also in
other's data sets\cite{giesenPtfluct,giesenAgfluct,giesenCufluct}
indicate the difficulty of measuring these barriers. This may be due to
the fact that the local environment (average number of kinks on the step
edge) can change significantly from point to point along the step and
also from measurement to measurement (as is shown quite eloquently in
Fig.~\ref{fig:tauvsT} for the two sets of measurements) as the
equilibration time on larger length-scales can be very large in
comparison to the measurement times. (See a more thorough discussion of
these experimental difficulties in M. Giesen's review
Ref.~[\onlinecite{01giesenrev}].) These measurements and especially the
prefactors may be sensitive to these details of the system
configuration, while other parameters seem to be more robust such as the
dynamical exponents.

In summary, using VT-STM we measured the fundamental step edge effective energy barrier characterizing the transport properties of the Pb(111) surface steps.  We mimicked this system using KMC simulations with realistic energy barriers in the framework of the nearest neighbor lattice gas model. The experimental and the simulational data agree reasonably well for low temperature so that we can deduce an energy barrier that shows the right trend for the barriers measured previously by other groups for Pt(111),\cite{giesenPtfluct} Ag(111),\cite{giesenAgfluct} and Cu(111).\cite{giesenCufluct}

The use of MC simulations apparently helps greatly to sort out the right energy barriers and attempt frequencies from the strongly scattering experimental findings and places these findings on firm footings as it makes the trends in the data more clear. However, the comparison of the measurement and the MC also reveals the range of applicability of simple models, like the one used in this study with nearest neighbor interactions and (short) nearest neighbor jumps. While it applies to low temperatures, the higher temperature parameter range apparently requires more complex models with possibly longer-range interactions and/or longer adatom jumps or other complicated atomic moves.

\section*{APPENDIX: Detailed-balance Considerations}

To describe equilibrium thermodynamic properties accurately, Monte Carlo algorithms should satisfy detailed balance, which insures that the equilibrium distribution is a steady-state solution.  Likewise, detailed balance is deemed crucial in simulations of systems near equilibrium, such as weakly driven systems.  (For systems far from equilibrium, such as kinetically-limited growth, algorithms that violate detailed balance by excluding highly improbable events can still give satisfactory results.)  In this appendix we illustrate explicitly that either of the break schemes satisfies detailed balance. 

We consider two different ways of performing a simple hopping process as depicted in Fig.~\ref{fig:microstates}.  In process $A\rightarrow B \rightarrow C$ an adatom hops from site $A$ to $B$ and then $C$ while in $A\rightarrow C$ it hops directly from $A$ to C. In both cases all the other adatoms remain fixed. $A$, $B$, and $C$ can also serve to denote the microstates with the corresponding adatom configurations and energies as in Fig.\ \ref{fig:elandscape}. The energy difference of two states $\Delta E_b$ is the difference between the energy barriers of the forward ($A\rightarrow B$) and backward ($B\rightarrow A$) moves between the states (e.g.\ $\Delta E_b^{A\rightarrow B}=E_b^{023}-E_b^{124}$ in the EAM barrier notation or $\Delta E_b^{A\rightarrow B}=E_b^{(0)}-E_b^{(1)}$ in the break-3 notation) for the $A\rightarrow B$ move. With a proper scheme for approximating the barriers, detailed balance is automatically satisfied. Following through this example for $A\rightarrow B$ and $B\rightarrow C$ compared to $A\rightarrow C$, the two paths lead to the same energy difference, $E_{TD}+ 2E_{NN}$, between microstates $A$ and $C$. 
\begin{figure}[t]
\includegraphics*[trim= 0 0 -40 0 ,
angle=270,scale=.5]{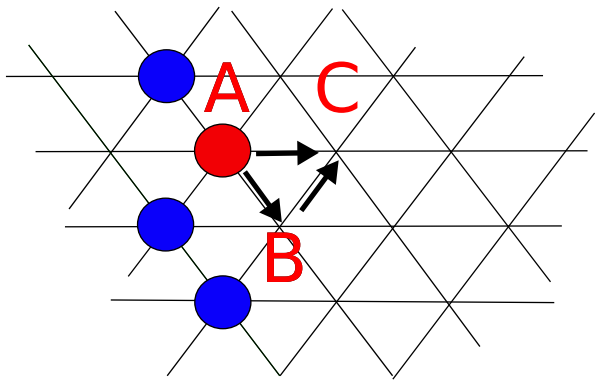}
\caption{ (Color online) Three microstates and the two processes $A \rightarrow B \rightarrow C$ and $A \rightarrow C$ discussed in the text to show how detailed balance is satisfied for the bond-breaking energy barriers and fails when simply using the NN EAM barriers.}
\label{fig:microstates}
\end{figure} 
However, doing the same with the EAM barriers the first and the second processes give different energy difference between $A$ and C: from Table \ref{tb:energybarriers} we see that the sum -32 + 225 = 193 meV (for $A\rightarrow B$ and $B\rightarrow C$ processes) differs significantly from 235 meV (for $A\rightarrow C$), the difference being 42 meV or about 20\%.  Thus, the NN EAM barriers do not satisfy detailed balance, as mentioned earlier. (When NNN's are appropriately taken into account, we have confirmed that detailed balance is precisely satisfied.  Thus, the energy landscape in EAM is well-defined.  The subtleties arise when reducing the parameter space to a computationally tractable size.) 

On very general grounds, bond breaking schemes satisfy detailed balance: the energy difference between states $i$ and $j$: 
\begin{equation}
E_b^{i\rightarrow j}-E_b^{j\rightarrow i }=(n_i-n_j) E_{\rm NN}
\label{eq:ijji}
\end{equation}
\noindent where $n_i$ and $n_j$ are the number of occupied nearest-neighbor sites of sites $i$ and $j$, respectively. Eq.~(\ref{eq:ijji}) is a straightforward consequence of the break-5 scheme. Moreover, it is also valid for the break-3 case---in which the bridge sites 2 and 2$^\prime$ are neglected in $n_i$ and $n_j$---since these two sites are neighbors of both sites $i$ and $j$; hence, they do not contribute to the \textit{difference} of $n_i$ and $n_j$.
\begin{figure}[t]
\includegraphics[angle=-90,width=\columnwidth]{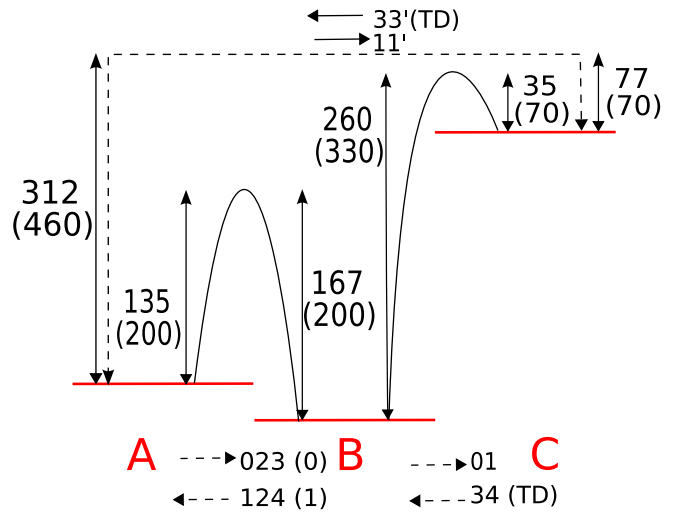}
\caption{\label{fig:elandscape} (Color online) Energy landscape for $A\leftrightarrow B\leftrightarrow C$ and $A\leftrightarrow C$ processes in Fig.~\ref{fig:microstates}. The processes at the bottom and the top are indicated by the occupied NN sites  around the hop (while in the parentheses we indicate the occupied sites relevant to the brake-3 model). The numbers next to the arrows are the EAM energy barriers in meV (in parentheses the corresponding break-3 barriers) taken from column four in Table \ref{tb:energybarriers} (and column two for the break-3 barriers).}
\end{figure} 

The energy difference in the situation when sites $i$ and $j$ are not neighbors but farther apart is still correctly given by Eq.~(\ref{eq:ijji}): the path between the initial and final positions can be decomposed into a (non-unique) sequence of NN hops.  Then the energy of each intermediate position is both added and subtracted, cancelling out, so that difference of the $n$'s of the initial and the final positions survive. Finally, since any configuration can be obtained from any other by moving adatoms one by one, every configuration has a uniquely defined energy. Hence, the energy barriers form a ``consistent" set. Writing the rate of move $i\rightarrow j$ as $r_{ij}=\nu_D \exp(-\beta E_b^{i\rightarrow j})$, where $E_b^{i\rightarrow j}$ is defined in Eq.\ (\ref{eq:Eb}) and $\nu_D$ is the Debye frequency, we see immediately that detailed balance is satisfied because the rates in the ratio
\begin{equation}
\frac{r_{ij}}{r_{ji}}={\rm e}^{-\beta(E_b^{i\rightarrow j}-E_b^{j\rightarrow i})}={\rm e}^{-\beta(n_i-n_j)E_{\rm NN}}
\end{equation}
are based on these energy levels of the configurations. This consistency due to the uniquely defined energies of the states characteristic of the bond-breaking schemes and also of the Kawasaki scheme is missing in the NN set of EAM numbers.  

\section*{Acknowledgements}
Work at the University of Maryland was supported by NSF-funded MRSEC there, via NSF Grant DMR 05-20471 and DOE-CMSN DEFG0205ER46227. We thank Dr. Bill Cullen for comments on the manuscript.

\newpage




\begin{thebibliography}{}
\bibitem{loT} The low temperature approximation is rather good in the temperature range in which we did our experiments and simulations; there is only 1.5\% difference between the exact Akustu formulas \onlinecite{akutsuakutsu} and the approximation.
\bibitem{type} The barriers are called Kawasaki-type rather than Kawasaki because $E_{TD}$ is included in the energy barrier in addition to the energy difference between final and initial states.  See Ref.~\onlinecite{NewBark99}.
\end{thebibliography}
\end{document}